# Dislocation-mediated and twinning-induced plasticity of CoCrFeMnNi in varying tribological loading scenarios

Antje Dollmann[1,2], Alexander Kauffmann[1], Martin Heilmaier[1], Aditya Srinivasan Tirunilai[1], Lakshmi Sravani Mantha[3,4], Christian Kübel[3,4,5], Stefan J. Eder[6,7], Johannes Schneider[1,2], Christian Greiner[*1,2]

[1] Institute for Applied Materials (IAM), Karlsruhe Institute of Technology (KIT), Kaiserstraße 12, 76131 Karlsruhe, Germany
[2] MicroTribology Center (µTC), Straße am Forum 5, 76131 Karlsruhe, Germany
[3] Institute of Nanotechnology (INT), KIT, 76344 Eggenstein-Leopldshafen, Germany
[4] KIT-TUD-Joint Research Laboratory Nanomaterials, Technical University Darmstadt (TUD), 64287 Darmstadt, Germany
[5] Karlsruhe Nano Micro Facility (KNMF), KIT, 76344 Eggenstein-Leopoldshafen, Germany
[6] AC2T research GmbH, Viktor-Kaplan-Straße 2/C, 2700, Wiener Neustadt, Austria
[7] TU Wien, Institute of Engineering Design and Product Development, Lehárgasse 6, Objekt 7, 1060, Vienna, Austria

*To whom correspondence should be addressed: greiner@kit.edu

Abstract

Coarse-grained, metallic materials undergo microstructure refinement during tribological loading. This in turn results in changing tribological properties, so understanding deformation under tribological load is mandatory when designing tribological systems. Single-trace experiments were conducted to understand the initiation of deformation mechanisms acting in various tribological systems. The main scope of this work was to investigate the influence of normal and friction forces as well as crystal orientations on the dominating deformation mechanism in a face-centred cubic concentrated solid solution. While varying the normal force is easily realised, varying friction forces were achieved by using several counter body materials paired against CoCrFeMnNi. The subsurface deformation layer was either mediated through dislocation slip or twinning, depending on the grain orientation and on the tribological system. A layer dominated by dislocation-based deformation is characterised by lattice rotation, the formation of a dislocation trace line or subgrain formation. Such behaviour is observed for tribological systems with a low friction coefficient. For systems dominated by deformation twinning, three types of twin appearance were observed: small twins interacting with the surface, large twins and grains with two active twin systems. Two different twinning mechanisms are discussed as responsible for these characteristics.

**Graphical abstract**

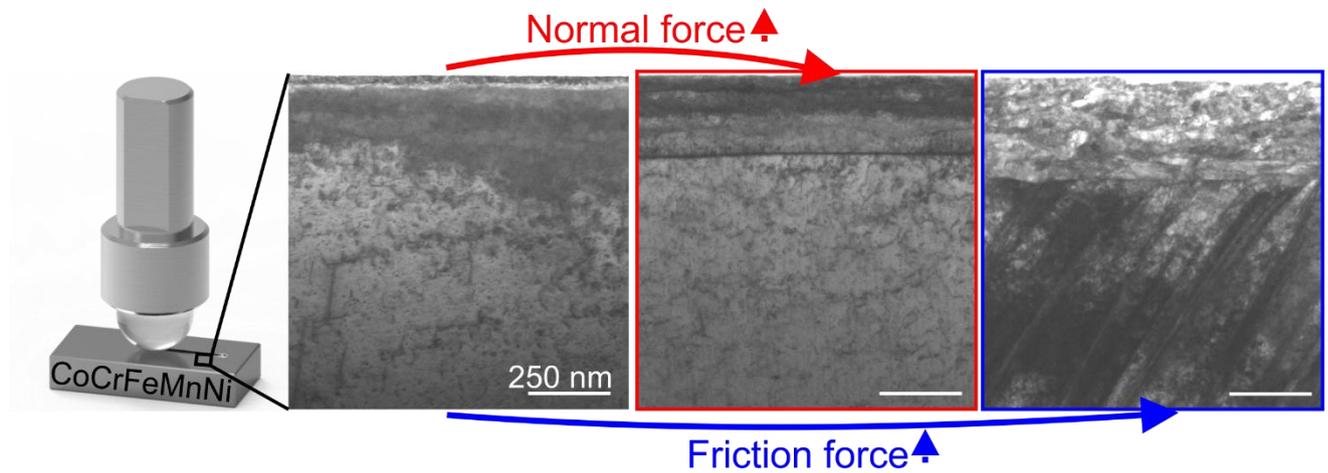

Introduction

Previous research has shown that for metals in a tribological contact, a feedback loop exists between the friction coefficient, the stress state and the microstructure [1]. During a tribological experiment, the friction coefficient and the normal force determine the stress state that leads to a microstructural evolution in the subsurface area [2,3]. In turn, the resulting deformation layer leads to a change in the friction coefficient. So far, various microstructural characteristics have been observed in coarse grained, face-centred cubic (fcc) metallic materials under tribological load, such as a dislocation trace line [4,5], subgrain formation [2,6], deformation twinning [7,8], band-like patterns [9–11] and nanocrystalline grains [7,12].

In essence, a tribological system features two forces controlling the microstructural evolution: first, the normal force, which can be easily adjusted and second, the friction force, which is a function of the tribological system itself. Further parameters determining the deformation layer are cycle number [4,13,14], sliding velocity [15][16] and temperature [16].

The quinary single phase high-entropy alloy (HEA) CoCrFeMnNi [17], exhibiting a medium stacking fault energy, was chosen as the material of interest because it covers dislocation-mediated microstructures at low strains and twin-induced plasticity at higher strains, although this knowledge stems from uniaxial load experiments [18,19]. This deformation behaviour makes it possible to assign the microstructures of CoCrFeMnNi that evolve under triblogical loading to distinct strain regimes under uniaxial load. To the authors' best knowledge, only three other research groups have investigated the tribological properties of single phase CoCrFeMnNi so far. Ayyagari *et al.* analysed the wear behaviour in dry and marine environments [20]. Joseph *et al.* [21] investigated the friction and wear behaviour at temperatures ranging from room temperature up to 900°C. Jones *et al.* examined the interplay between grain refinement and the friction and wear behaviour after several thousand sliding cycles [12]. The two normal loads they applied resulted in different grain sizes in the subsurface area [12]. In our earlier work, we analysed the microstructural evolution with increasing cycle number with a focus on the onset of microstructural changes at high friction coefficients [7]. In contrast, the influence of the friction and normal force is investigated after a single-trace experiment in this manuscript. Besides that, the pronounced adhesive tendency of CoCrFeMnNi enables us to cover friction coefficients between 0.1 and 1 by using different counter body materials and environments. Furthermore, the influence of an increased normal load is investigated.

The polycrystallinity of CoCrFeMnNi makes it necessary to take the crystal orientation into account. It is known from uniaxial tests that the initial grain orientation is decisive for the occurrence of deformation twinning [22,23]. In contrast to high SFE materials [24–26], there is only little data available on medium SFE materials [7,13,27]. For example, nanoscratch tests in austenitic steel have demonstrated the influence of the initial crystal direction parallel to the normal direction on the formation of slip traces next to the wear track, pile-up and the wear track depth [27]. Hence, with the study of polycrystalline CoCrFeMnNi, it is a further goal of this manuscript to draw conclusions about the relation between the active deformation mechanism and the crystal orientation under tribological load.

Materials and Methods

CoCrFeMnNi was manufactured via arc-melting, homogenised, cold-rolled and heat treated as described elsewhere [7]. An average grain size of about 40 µm (excluding recrystallisation twins) was measured by the linear intercept method. The chemical composition was determined via inductively coupled plasma optical emission spectrometry (ICP-OES) to 15.2 at-.% Co, 21.7 at-.% Cr, 21.2 at-.% Fe, 21.2 at-.% Mn and 20.1 at-.% Ni (accuracy of 0.1 at-.%). To ensure a defect-free surface, the samples were ground up to a grid of P4000, polished with diamond suspensions of 3 µm and 1 µm (Cloeren Technology GmbH, Wegberg, Germany) for at least eight minutes each, and finally electropolished with an electrolyte composed of perchloric acid and methanol in a ratio of 1:9. Images of the defect-free surface are given in Ref. [7]. A surface roughness of $S_a = (80 \pm 10)$ nm was achieved with the described procedure. The measurements were carried out interferometrically with a PLu neox instrument (Sensofar, Barcelona, Spain).

The tribological experiments were conducted with a custom-built tribometer encapsulated in a climate-controlled chamber. Counter body spheres with a diameter of 10 mm made of sapphire (Saphirwerk, Switzerland), SiC (hightech ceram, Dr. Steinmann + Partner GmbH, Germany) and $Si_3N_4$ (Saint-Gobain Ceramic Materials GmbH, Germany) were used. The surface roughnesses $S_a$ are $(14 \pm 2)$ nm for sapphire, $(20 \pm 4)$ nm for $Si_3N_4$ and $(60 \pm 12)$ nm for SiC, interferometrically measured. All experiments were conducted at room temperature in single trace mode at a sliding velocity of 0.5 mm/s and a stroke length of 12 mm. Experiments were performed in air with (50 ± 3)% relative humidity (RH) and a normal load of 2 N with $Si_3N_4$, SiC and sapphire spheres as counter bodies. Further experiments were run in dry $N_2$ atmosphere with a normal load of 2 N and SiC and sapphire spheres as counter bodies. An additional experiment was conducted at a normal load of 5 N in air with (50 ± 3)% RH and a SiC sphere. The wear tracks were analysed using a dual beam focused ion beam (FIB), scanning electron microscope (Helios NanoLab™ DualBeam™ 650, ThermoFisher, Hillsboro, USA). TEM foils were cut parallel to the sliding direction in the middle of the wear track in a process with negligible ion beam damage [28]. Images were taken in STEM mode in the SEM. Additional analyses were conducted using on-axis transmission Kikuchi diffraction (TKD). The scans were performed with an acceleration voltage of 30 kV, a current of 0.8 nA and a step size of 5 nm. The patterns were recorded by an e-flash$^{HD}$ detector with the Optimus$^{TM}$ TKD head and analysed using Esprit 2.1 (all by Bruker, Billerica, USA). These data were further analysed with the Matlab toolbox MTEX [29].

High-resolution transmission electron microscopy (HRTEM) images were obtained using an (image) aberration-corrected Titan 80-300 (ThermoFisher Scientific, Hillsboro, USA) operated at 300 kV and equipped with a Ultrascan 1000 CCD camera (Gatan, Pleasanton, USA). These images are average background subtraction filtered (ABSF) in DigitalMicrograph® [30].

Results

The various tribological systems investigated here were selected to result in varying friction coefficients (µ). The friction coefficients as a function of sliding distance for the single trace experiments are given in Figure 1 for all six systems. In Figure 2, the cross-sectional STEM images of each tribologically loaded CoCrFeMnNi system are given. A colour was assigned to each tribological system. If two grains of a tribological system were analysed, different shades of the colour were used. The system SiC/2N/air is colour-coded in green, SiC/2N/$N_2$ in blue, sapphire/2N/$N_2$ in purple, sapphire/2N/air reported in [7] in orange, $Si_3N_4$/2N/air in black and SiC/5N/air in red. This colour-coding for the considered tribological systems and grains is used throughout the entire manuscript. A grey vertical line in Figure 1 at a sliding distance of 6 mm marks the STEM foil lift-out position. The friction coefficient at this position is therefore the one determining the microstructures observed using (S)TEM.

SiC spheres paired with CoCrFeMnNi result in the lowest observed friction coefficient, independent of atmosphere or applied normal load, and maintained an almost constant friction coefficient with sliding distance. The change from air with 50% RH to dry $N_2$ causes a small decrease in the friction coefficient at a normal load of 2 N. The increase in normal load from 2 N to 5 N in air with 50% RH results in a similar friction coefficient. The tribological systems sapphire/2N/$N_2$ and $Si_3N_4$/2N/air exhibit a varying friction coefficient with increasing sliding distance. Interestingly, both tribological systems have the same friction coefficient at a sliding distance of 6 mm. The highest friction coefficient was measured for the system sapphire/2N/air [7].

The investigated tribological systems also differ in the occurrence of material transfer. All wear track and sphere images are given in the supplement Figure S1. No material transfer is observed in any experiments with a SiC sphere as counter body, whereas the wear tracks of the other tribological systems exhibit flake formation and material transfer to the counter body.

STEM images of the subsurface deformation layer in CoCrFeMnNi for all experiments are given in Figure 2. Images of two grains of the experiment with SiC/2N/air are presented in Figure 2a+b to investigate the initial grain orientation. The microstructure in Figure 2a is characterised by a bright horizontal line at a depth of $(26 \pm 5)$ nm. In contrast, nearly vertical line-type features occur down to a depth of $(163 \pm 49)$ nm in Figure 2b. The bands are slightly tilted in sliding direction (SD). The subsurface deformation layer for the experiment with SiC/2N/$N_2$ is given in Figure 2c. The STEM foil includes two grains, and the initial grain boundary is marked in the image. The microstructure is dominated by a line starting at the surface in the left grain, crossing the initial grain boundary and proceeding into the right grain down to a depth of 290 nm. A horizontal line is observed closer to the surface marked by a black arrow. In the tribological system $Si_3N_4$/2N/air, two sets of line-type characteristics are observed, as seen in Figure 2d. Here, one line is again tilted in SD and the other opposite to SD, both marked in white colour. The microstructure after the experiment with sapphire/2N/$N_2$ in Figure 2e consists of a nanocrystalline layer directly in the sub-surface area and line-type characteristics tilted in SD. A horizontal line at a depth of $(255 \pm 15)$ nm is observed for SiC/5N/air in Figure 2f, with a grey gradient between the line and the surface.

The described deformation layers are further analysed by transmission Kikuchi diffraction (TKD) and HRTEM. In Figure 3a to f, TKD measurements of the STEM images in Figure 2a+b of the experiment SiC/2N/air are presented. Both measurements are colour-coded in sliding (SD), normal (ND) and transverse direction (TD). Furthermore, MTEX was employed to detect the different types of grain boundaries. Grain boundaries with a small-angle boundary character (SAGB) (3° to 15°) are coloured in green, with a high-angle character (HAGB) (>15°) in red and twins (Σ3) in blue. The horizontal line in Figure 2a was identified as a SAGB in Figure 3a to c. A colour-gradient is seen beneath the SAGB with increasing distance from the surface. The line-type characteristics in Figure 2b could not be resolved in Figure 3d to f, instead a colour-gradient is observed.

The crystal orientations between the white lines in Figure 3a+d are plotted in inverse pole figures (IPFs) in Figure 3g to h in SD, ND and TD, respectively. The grain orientations farthest away from the surface are marked by a circle and are interpreted as the initial grain orientations. The orientations at the surface are marked by arrow heads. The dashed line represents the data points of the grain in Figure 3a, and the solid line the data points of the grain in Figure 3d. The initial crystal directions parallel to SD are both close to the [111] pole in Figure 3g. Both crystal directions move along a bisector with decreasing distance to the surface. The crystal direction parallel to SD of the grain in Figure 3a moves along the [001]-[111] bisector, leaving the bisector close to a [113] direction, rotating towards the [001]-[011] bisector. The crystal direction of the grain in Figure 3d moves along the [011]-[111] bisector in Figure 3g. The crystal direction parallel to ND in Figure 3h of the grain in Figure 3b is reflected twice at two different bisectors, whereby the crystal direction parallel to ND of the grain in Figure 3e rotates from the centre of the IPF into the direction of the [001] pole. The crystal direction parallel to TD in Figure 3i changes only a little for both grains. The misorienation was calculated in reference to the initial grain orientation. Misorientation profiles were extracted along the arrows in Figure 3a+d and plotted in Figure 3j. The one of the grain in Figure 3a exhibits a discontinuity smaller than 10° and has a maximum misorientation at the surface of 35°. In contrast, the profile of the grain in Figure 3d is continuous and exhibits a maximum misorientation of 10°. Both grains have in common that the increase in misorientation starts at a depth of around 0.5 µm.

The line-type features in Figure 2b were further investigated using HRTEM. The result of this analysis is given in Figure 4. The sample surface is marked by a dashed line. In three different regions with increasing distance to the surface, fast Fourier transformation (FFT) analyses were performed in order to investigate the possible occurrence of twinning. The blue dashed line in the FFT images connects the diffraction spots for the matrix, while the red dashed line connects the diffraction spots of the twins. In the regions closer to the surface, the twin is unambiguously detected, which is not the case for the area farthest from the surface. The twinned area is thicker close to the surface than in the bulk also seen by surface steps in the twin boundary. Furthermore, the twin itself has not a perfect crystal structure.

The TKD measurements of the experiments with SiC/2N/$N_2$ (Figure 2c) and SiC/5N/air (Figure 2f) are presented in Figure 5. In all grains, the sharp line demarking the deformation layer is identified as a SAGB. A colour-gradient is observed above the SAGB in all three grains. Additionally, SAGBs are detected directly beneath the surface. The same analysing procedure described for Figure 3 was used to generate the IPFs (Figure S3) and misorientation profiles. The misorientation profiles of all three grains in Figure 5g have in common that they exhibit discontinuities and a maximum misorientation of around 55°. The depth of the discontinuity of the right grain of the SiC/2N/$N_2$ experiment and the grain in the SiC/5N/air experiment is nearly the same. In contrast, the discontinuity in the left grain of the SiC/2N/$N_2$ experiment is closer to the surface. The magnitude of the discontinuity is the largest for the SiC/5N/air experiment.

The TKD measurements with superimposed grain boundary analyses of the experiments with sapphire/2N/$N_2$ (Figure 2e) and $Si_3N_4$/2N/air (Figure 2d) are given in Figure 6. The microstructure of the sapphire/2N/$N_2$ experiment (Figure 6a to c) is characterised by nanocrystalline grains in the subsurface area as well as by twins. The two regions are separated from each other by a SAGB. The nanocrystalline grains become coarser with increasing distance to the surface. The twins exhibit lenticular shape, hence, being thinner at their ends than in the centre. The subsurface region of the experiment $Si_3N_4$/2N/air (Figure 6d to e) exhibits a layer with grains separated by SAGBs beneath the surface, and with increasing distance to the surface two twin systems are detected. The line-type characteristics (see Figure 2d) exhibits twin and high-angle grain boundaries.

Discussion

This manuscript investigates the influence of a range of tribological systems on the frictional properties and especially on the subsurface microstructure in order to answer the following questions: How do friction and normal load change the tribology-induced microstructural evolution? Which are the characteristics and formation mechanisms of the observed microstructures developing under varying tribological loads? Which role does the initial grain orientation play?

**Variation of friction forces**

The tribologically imposed stress states of the materials constituting the contact is a function of the friction coefficient [1,31]. To investigate the effect of the stress state on the microstructural evolution, the tribological systems were varied to achieve varying friction coefficients, see Figure 1. Steady-state friction coefficients are already published in literature for several tribological systems containing CoCrFeMnNi [12,20,21], which are similar to the friction coefficients in the present study. The obtained variation of the friction coefficients stems from various contributions as briefly discussed below.

The work of adhesion is considered in the first place, which comprises the surface free energies of the two materials and their interface [32]. Neither the surface free energy of CoCrFeMnNi nor the ones of the interfaces are published in literature to the best of the authors' knowledge. The values of the surface free energy of the counter body materials [33] do not directly correlate with the measured friction coefficients. No final conclusion can therefore be drawn on the influence of the work of adhesion on the friction coefficient in the chosen tribological systems. However, wear tracks with a low friction coefficient have a scratched appearance, and flake formation was observed in wear tracks with a high friction coefficient (Figure S1). Flake formation is indicative of adhesive wear [32]. These factors therefore provide a qualitative indication of different adhesive forces within the chosen tribological systems, but the adhesive forces will not be further analysed in the current manuscript.

The second parameter is the surface roughness of the different counter body materials. It was proposed in literature [34–36] that with an increasing surface roughness the contact area decreases, which can result in a lower friction coefficient. This tendency was confirmed in our experiments.

The environmental changes from air with 50 %RH to dry $N_2$ resulted in a lower friction coefficient. Surface softening was reported for sapphire with increasing humidity [37][38] which can facilitate shearing of the surface and lead to an increased contact area [39]. These two effects might therefore cancel each other out [39].

Three aspects, namely adhesive force, surface roughness and surface softening, were considered to explain the varying friction coefficients. So far, it is unclear to which extent they contribute to the friction forces.

**Influence of initial grain orientation**

The initial grain orientation can strongly influence the deformation mechanism, especially in the case of deformation twinning [22,23]. The influence of the initial grain orientation was investigated by comparing two subsurface deformation layers for the same tribological system and, therefore, also the same applied surface stress state.

The tribological system in focus was chosen to be SiC/2N/air, as it does not exhibit material transfer and the friction coefficient is nearly constant over the entire sliding distance (see Figure 1a). The two grains in question were presented in Figure 2a+b and Figure 3, demonstrating their different microstructural evolution. The thickness of the tribologically induced subsurface layer is small for both grains, so the orientation underneath this deformed layer was interpreted as the initial grain orientation (see TKD results in Figure 3).

The SAGB identified in Figure 3a has been described as a dislocation trace line (DTL) in earlier publications and is believed to be caused by dislocation self-organisation [4,5,13]. Until now, the DTL has never been observed as close to the surface as in Figure 3a–c. This can be an effect of the grain orientation and differing mechanical properties of CoCrFeMnNi compared to Cu, as previously DTL formation was studied for Cu and its alloys [4,13,14]. The DTL in CoCrFeMnNi will be further discussed in the subsection 'Dislocation trace line and crystal rotation'. The line-type features in Figure 2b were clearly identified as twins using FFT in Figure 4. Therefore, there is a need for understanding why plastic deformation is mediated by dislocation slip in one grain and twinning-induced in the other; under otherwise identical tribological conditions. The initial grain orientation as well as the type of applied load is decisive for deformation twinning [23]. The loading was the same for both grains. The initial crystal directions of the two grains are quite close to each other parallel to SD and TD (Figure 3g and i). Thus, it is evident that the initial crystal direction parallel to ND under the given loading parameters is decisive for the dominating deformation mechanism. This is unexpected, as it was demonstrated in a previous study that the initial grain direction parallel to SD is decisive for the occurrence of deformation twinning [7]. The differences between the published results in [7] and the results in Figure 3 are the selected counter body materials, resulting in a significantly reduced friction coefficient and no material transfer here and in a higher friction coefficient and material transfer in [7]. The lack of material transfer results in lower adhesive forces, which in turn decrease the tensile stresses parallel to SD at the trailing edge of the counter body. Further details about the twinning mechanism under tribological load are discussed in the section 'Twinning under varying loading conditions'.

The increase in misorientation in Figure 3j is caused by dislocation motion. In both grains this increase starts at 0.5 µm, which shows that the critical resolved shear stress (CRSS) for dislocation glide is reached at a similar depth. A difference between the two grain orientations can be observed in the maximum misorientation from the bulk to the surface which, is approximately 25° higher in the grain with the DTL than in the grain with twins. The misorientation profiles (Figure 3j), however, only include the misorientation changes caused by dislocation motion. This indicates a difference in the dislocation activity. The discontinuity in misorientation corresponds to the SAGB in the grain shown in Figure 3a–c. The higher the misorientation, the more dislocations are stored in the lattice. The formation of a SAGB leads to a reduction in the stored lattice energy, therefore the discontinuity could be favoured [40].

The two investigated grains under the same tribological loading reveal that not only the stress state, but also the initial grain orientation can be decisive for the dominating deformation mechanism.

**Influence of a higher normal load**

For medium stacking fault energy materials like CoCrFeMnNi, increasing the stress under uniaxial loading increases the probability for the material to twin. We therefore expected to observe more and larger twins at a normal load of 5 N than at a normal load of 2 N (Figure 4). However, the microstructures in Figure 2f and Figure 5d–f for the SiC/5N/air experiments do not show any evidence of twinning. In contrast, we found DTL formation [4,5,13]. The same was observed for a deformation layer within another grain in the same experiment (see Figure S2). The differences between the subsurface microstructures in these two grains are the DTL depths and the grain boundary type of the DTL, which is a SAGB (Figure 5d) in one grain and a HAGB in the other (Figure S2). Additionally, the area between the DTL and the surface is dominated by subgrains in Figure S2 and does not exhibit a colour-gradient as observed in Figure 5d. The microstructural features are formed in both grains by dislocation glide and self-organisation. The same was observed and already discussed for the grain in Figure 2a. Both experiments showed a similar friction coefficient, which also means that the friction force is higher with normal load. This is why the influence of the increased friction force on the DTL depth cannot be separated from the increased normal load. The higher normal load results in a larger misorientation difference at the discontinuities and a higher maximum misorientation at the surface in the misorientation profiles (Figure 3j and Figure 5g). This can be interpreted as stronger dislocation activity with higher normal forces. Some insights into crystal rotation are given in the next section. Contrary to expectations, an increase in normal load did not increase the twinning probability, but rather the amount of dislocation activation and motion.

**Dislocation trace line and crystal rotation**

Four grains shown in Figure 2 exhibit a DTL [4,5,41] even though their grain orientations and tribological loading differ. For this reason, the DTL and its characteristics will be discussed in detail. Dislocation self-organisation is presumed to be the responsible mechanism for DTL formation [4,41]. CoCrFeMnNi has a medium SFE [42] resulting in mainly planar slip. It was reported in previous work that these planar slip materials exhibit band-like patterns in the subsurface microstructure after tribological loading [7,9,11,13]. However, exceeding a critical dislocation density might have led here to dislocation self-organisation forming the DTL.

The depth of the DTL cannot be correlated with the applied load because the DTL for the experiment SiC/2N/$N_2$ with the lowest friction coefficient is located at the largest distance to the surface (Figure 2c). This seems to be contrary to results presented in reference [43], but if the lower and not the upper DTL is considered in [43], the results agree well. The crystal orientation is apparently a key parameter concerning the DTL depth. While here the depth of the DTL seems to be independent of the normal load, the misorientation discontinuity at the DTL with a normal load of 5 N is the highest. From this we can conclude that the loading condition can influence the degree of the misorientation discontinuity at the DTL. All observed DTLs have in common that their crystallographic directions parallel to SD and ND are close to a bisector. Two slip systems are active under uniaxial loading if the loading axis of a single crystal is located on a bisector, while only one is active in the centre of the IPF. It is expected that the number of active glide systems also increases under tribological loading if a crystal direction parallel to a SD or ND is on a bisector, but the number of activated slip systems is unclear. If uniaxial loads are considered parallel to the sample axis, this position close to a bisector could mean that the increased number of activated slip systems is able to accommodate the applied strain more easily, resulting in a misorientation discontinuity.

Furthermore, the DTL is only observed in systems with a low friction coefficient and without flake formation in the wear track. The stress state beneath the moving sphere changes with varying friction coefficient [31]. At the same time, the friction force can be interpreted as the sum of forces arising by ploughing and by adhesion [44], with the two being difficult to separate. Therefore, the occurrence of material transfer can further alter the stress state, and to our knowledge, no adequate stress field model exists to cover this behaviour.

As mentioned before, the initial crystal orientation seems to play a key role on the microstructural evolution as in uniaxial tensile experiments. Here, the DTL forms if crystal directions parallel to SD and ND are located in the [012]-[011]-[111]-[113] region of the IPF, see Figure 3 and S3. Based on the present results in comparison to Ref. [7], it seems that for a low friction coefficient with no material transfer the crystal orientation with respect to ND is decisive, while when considering a high friction coefficient with material transfer, the crystal orientation with respect to SD is important.

The crystal rotation during tribological loading is an important parameter and can influence the dominant deformation mechanism. The smallest rotation of a crystal direction parallel to a principal sample axis is observed for the crystal direction parallel to TD in the IPFs in Figure 3i and Figure S3c for experiments with a low friction coefficient. This means that up to a certain stress level, TD is favoured as rotation axis. The current data might allow the interpretation that such behaviour is not limited to a specific initial crystallographic orientation. The crystal rotation around TD was also observed in other studies [4,10,43]. The directions parallel to SD and ND change through crystal rotation, as seen in Figure S3a+b. It is interesting to note that the movement of the crystallographic directions parallel to SD of the grain in the SiC/5N/air experiment and the left grain of the SiC/2N/N$_2$ experiment are similar, with a nearly identical initial grain direction parallel to SD (Figure S3a). The same was observed for the grain in the SiC/5N/air experiment and the right grain in the SiC/2N/N$_2$ experiment in ND (Figure S3b). However, more data is needed to verify this and understand the observed texture evolution.

The crystal direction parallel to SD in the SiC/2N/air experiment (Figure 3g), coloured in dark green, first moves along the [001]-[111] bisector towards the [001] pole with decreasing distance to the surface until [113], where the crystal rotates towards the [001]-[011] bisector. This means that the given rotation changes its direction. This might be explained by a change in glide behaviour as the direction parallel to SD is close to [113] at the turning point. In uniaxial tension or compression experiments, the stacking fault width is independent of the stress state along the line between [012] and [113]. If this line is crossed under uniaxial loading, the stacking fault width either decreases or increases, depending on the applied stress [23]. Further tests are required to identify whether this also holds true for tribological experiments.

Summarising this section, dislocation mediated microstructures of CoCrFeMnNi are caused by low friction coefficients, and they favour the formation of a DTL. The crystal directions parallel to SD and ND next to a DTL were observed to be close to bisectors, increasing the amount of active slip systems. The crystal rotation is mainly about TD for low friction coefficients.

**Twinning**

Besides dislocation activity, also twins were observed. The occurrence of twins can be predicted for uniaxial loading when the crystal orientation is known. This is not the case for a tribological load because of the associated complex stress field. This leads to the following questions: Do the twins form and grow in the compressive or the tensile part of the stress field caused by the moving sphere? Can a direction of the sample coordinate system be identified along which the stresses lead to twinning? Which twinning mechanism is active? We are of the opinion that the data presented here may give first insights towards answering these questions.

Three different stages of twinning are observed: The twin in the SiC/2N/air experiment (Figure 2b) is small, and only one twin system was active. The twin interacts with the surface. The deformation layer in the experiment sapphire/2N/$N_2$ (Figure 2e) also shows only one active twin system. However, the twin itself is large and wide, and it does not have any connection to the nanocrystalline layer above. The grain in the Si$_3$N$_4$/2N/air experiment (Figure 2d) exhibits two activated twin systems under tribological loading similar. The twins interact with the SAGB in the near-surface region.

As mentioned in the previous section, the force necessary to activate twinning in the SiC/2N/air experiment (Figure 2b) is most likely a compression stress parallel to ND. The initial grain direction parallel to ND of the grain featuring twins is not within the [001]-[012]-[113] sector, which precludes twinning under uniaxial compression load. Assuming that the force parallel to ND leads to twinning, the observed crystal rotation towards the [001] pole is necessary for twin formation, as the rotation increases the probability of twinning. Since the twins are small and close to the surface, it is unclear if the Venables twinning mechanism [22] is active in such a limited volume. Another possibility is that the twins were formed by the slip of Shockley partial dislocations on adjacent {111} planes starting from the surface. This mechanism was observed in MD simulations for nanocrystalline grains [45] and thin films [46]. Further analyses of the MD simulation in [47] given in Figure S5 also show that twins form at the surface under tribological load. The HRTEM image in Figure 4 suggests the second possibility, as the twin is thicker at the surface. Shockley partial dislocations are nucleated at the surface, and some of them may glide over longer distances than others owing to local differences in the elemental distribution [48]. Steps at the twin boundary are clearly visible and support this hypothesis. In Ref. [16], twinning is observed in Cu under tribological load at high sliding speeds and at cryogenic temperatures, and it is supposed to start at the surface roughness sites as well. The morphology of the twin is also a further indicator for the above mentioned mechanism.

The twins generated in the experiment with sapphire/2N/N$_2$ and detected using TKD measurements presented in Figure 6a–c are of lenticular shape, which is typical of deformation twins in coarse-grained materials. This is an indicator for the Venables mechanism being active. The crystal directions of the matrix parallel to SD, ND and TD are in the [012]-[011]-[111]- [113] region of the IPF in Figure 6. Under these conditions, only tensile stresses parallel to SD can result in twinning [7,23]. Such necessary tensile stresses parallel to SD may very likely be caused by adhesive forces between the alloy and the sliding sphere [7]. In Ref. [7], it was proposed that twin fragmentation results in a nanocrystalline layer under tribological load. For the results presented here, the twins seem to end beneath the nanocrystalline layer. The transition between the twinned region and the nanocrystalline layer is a line nearly parallel to the surface and of a SAGB character. We speculate that first a DTL was formed at the depth of the transition between the nanocrystalline and the twinned region. Above the DTL, dislocation self-organisation and/or dynamic recrystallisation lead to the formation of the nanocrystalline layer. The distance between the DTL and the surface is too small for the Venables twinning mechanism to be active. The fact that some grains in the nanocrystalline layer are elongated in SD also strongly suggests twinning not to be active. Such an elongation is widely observed in Cu [2,6] and has been attributed to dislocation self-organisation.

In the systems discussed so far, only one active twin system was detected, while for $Si_3N_4$/2N/air (Figure 2d), two twin systems are present. In the near-surface region in Figure 6d–f some SAGBs are visible. The ends of the twins closer to the surface interact with the SAGB. It may be possible that Shockley partial dislocations nucleate at the SAGB and glide on adjacent {111} planes, forming the twins. This seems likely as the twins have their largest width at the intersection with the SAGB. Based on MD simulations, easier activation of twinning processes was observed at rough grain boundaries [45], as is the case for the SAGB in Figure 6d–f. Nevertheless, the Venables twinning mechanism cannot be ruled out completely. It is possible that the twins were nucleated in the bulk region and then interact with the SAGB. An estimation of the stress direction responsible for twinning is not straightforward. A compression stress parallel to TD may lead to the activation of both twin systems. It is, however, unclear where such a stress in the centre of the wear track along TD might originate from. No conclusions can be drawn whether stresses parallel to SD or ND mediate twinning. It seems to be favourable, for multiple twin systems to be active under tribological load, when the grain directions parallel to SD or ND, or both, are located near the centre of the IPF. Additionally, as mentioned earlier, part of the twin boundary exhibits HAGB characteristics, likely a result from the compounded effect of significant interaction between twin boundary and dislocations. This implies that in the $Si_3N_4$/2N/air case not just twin activity, but also twin-boundary–dislocation interaction is high. In addition, the tribological system $Si_3N_4$/2N/air with a different initial grain orientation results in the microstructure presented in Figure S4. Here, neither a DTL nor twins are detected. Instead, a colour-gradient in the TKD measurement is observed and interpreted as crystal rotation [49] caused by dislocation motion. This demonstrates the tremendous influence of the initial grain orientation on the dominant deformation mechanisms.

Based on this discussion, two different types of twins were identified: The first one seems to be caused by compression stresses along ND when Shockley partial dislocations nucleate at the surface and glide on adjacent {111} planes. These twins form under lower friction forces and in the absence of material transfer. They are also small in size and width, effectively revealed only via HRTEM and not even by TKD. The second type of twins might arise from the Venables mechanism and be caused by a tensile stress parallel to SD.

Conclusions

We investigated the influence of the friction and the normal force on the dominating deformation mechanisms in CoCrFeMnNi for tribological single-trace experiments. A range of friction forces resulted from selecting varying counter body materials and atmospheres. The use of a polycrystalline material allowed us to give first insights into the role of the initial grain orientation. The ensuing subsurface microstructures were investigated by applying STEM, TKD and HRTEM. The following conclusions can be drawn:

- The initial grain orientation has a strong influence on the dominating deformation mechanism.
- With the selected tribological systems, dislocation-mediated as well as twinning-induced plasticity microstructures were obtained.
- The responsible force direction and mechanism for deformation twins depends on the tribological loading. With low friction coefficients, twins are most likely formed by glide of partial dislocations from the surface on adjacent {111} planes. Such twins most probably form under compression parallel to ND. With higher friction coefficients, tensile stresses parallel to SD may lead to twins formed by the Venables mechanism.
- In materials with a medium stacking fault energy such as CoCrFeMnNi, a dislocation trace line (DTL) can be formed, which is unexpected based on the planar slip behaviour. An increased number of active slip systems is anticipated when the crystal directions parallel to SD and ND are close to a bisector next to a DTL. The loading condition most likely influences the degree of the misorientation discontinuity along the DTL.
- Twins are favoured for higher friction forces, and a DTL is preferential at higher normal loads.
- The crystal rotation occurs nearly perfectly around TD at low friction forces.

Acknowledgement

CG acknowledges support by the German Research Foundation (DFG) under Project GR 4174/5-1 as well as by the European Research Council (ERC) under Grant No. 771237, TriboKey. AK thanks financial support by German Research Foundation for grant no. KA 4631/1-1. SJE acknowledges the Austrian COMET-Program (Project K2 InTribology1, no. 872176) and the endowed professorship tribology at the TU Wien (grant no. WST3-F-5031370/001-2017). LM thanks the Deutscher Akademischer Austauschdient (DAAD) for a PhD scholarship. The authors acknowledge the chemical analysis by ICP-OES at the Institute for Applied Materials (IAM-AWP), Karlsruhe Institute of Technology (KIT) and the help in sample preparation by Dirk Seifert and Jens Freudenberger, both from IFW Dresden. Computational results were obtained using the Vienna Scientific Cluster (VSC).

Declaration of Competing Interest

The authors declare that they have no known competing financial interests or personal relationships that could have appeared to influence the work reported in this paper.

Data availability

The data that support the findings in this study are available under the link DOI10.5445/IR/100039657 and from the corresponding author upon request.

Figures

**Figure 1. Friction coefficients as a function of the sliding distance.** The position at a sliding distance of 6 mm is marked, as this is the TEM foil lift-out position. The data for sapphire/2N/air (orange) is taken from [7].

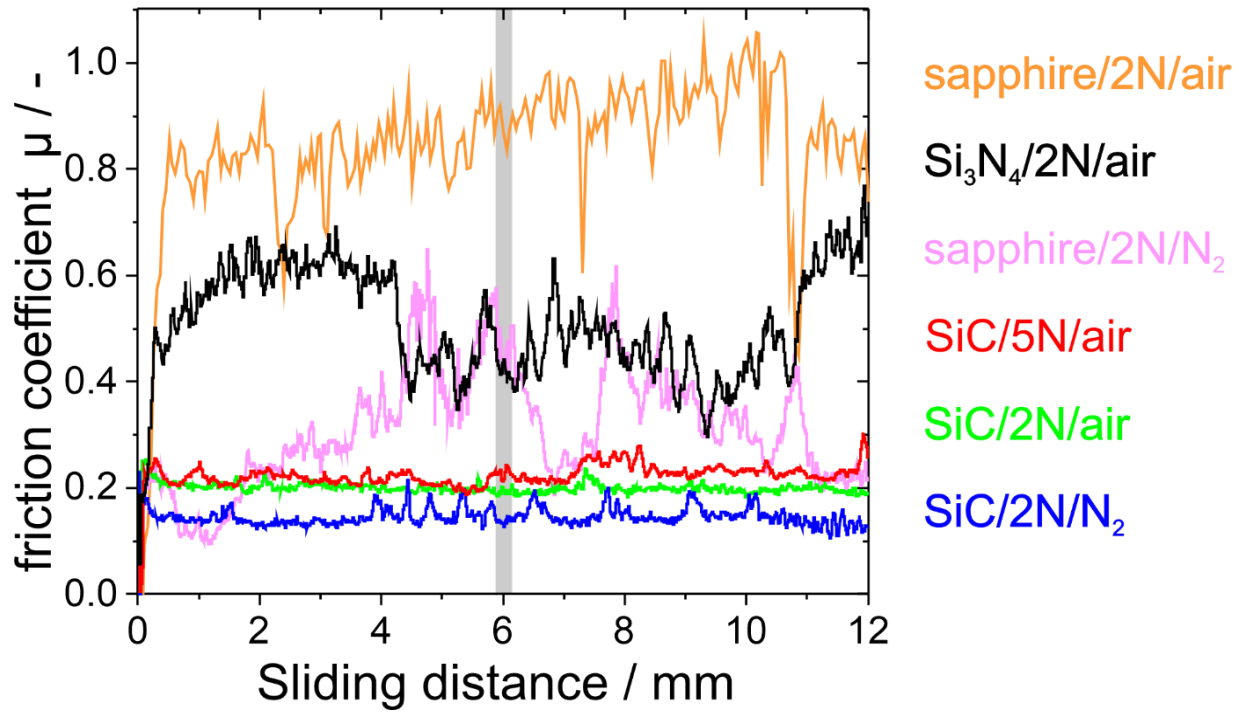

**Figure 2. STEM images of the various tribological systems.** a) and b) STEM images of the microstructures from the SiC/2N/air (green) experiment for two different grains. In b) a zoom-in of the near-surface area is marked by a black rectangle. c) STEM image of the microstructure from the SiC/2N/N$_2$ (blue) experiment, d) Si$_3$N$_4$/2N/air (black), e) sapphire/2N/N$_2$ (purple), and f) SiC/5N/air (red). All STEM images were cut in the middle of the wear track parallel to the sliding direction (SD). The SD is from left to right.

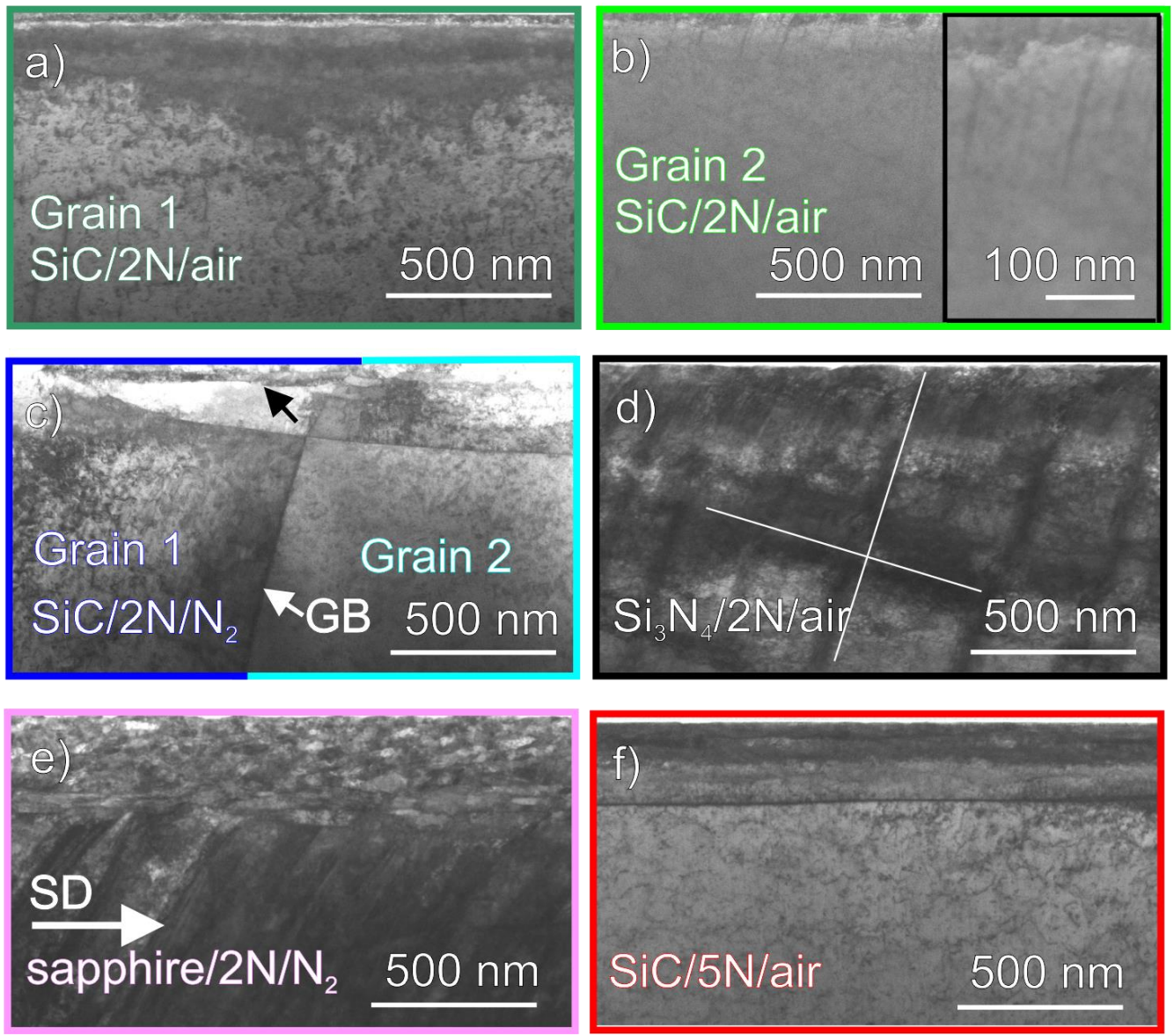

**Figure 3. Crystal orientation influence for the same applied stress state (SiC/2N/air).** a) to c) TKD measurement of the grain shown in Figure 2a, colour-coded in sliding (SD), normal (ND) and transverse direction (TD), respectively. d) to f) TKD measurement of the grain in Figure 2b, colour-coded in SD, ND and TD. The colour-coding is given in the inset in c). Small-angle grain boundaries (3-15°) are highlighted in green, high-angle grain boundaries (>15°) in red and twins (Σ3) in blue. The SD is from left to right. The orientation data between the white lines in a) and d) were plotted in g) to i) in inverse pole figures (IPFs) in SD, ND and TD, respectively. Dark green data points belong to the grain presented in a) to c) and in light green to d) to f). The dot marks the orientation in the bulk, and the arrowhead marks the orientation at the surface. The arrows within the IPFs point towards the surface. j) Misorientation profile along the arrows in a) and d). The distance 0 µm marks the surface.

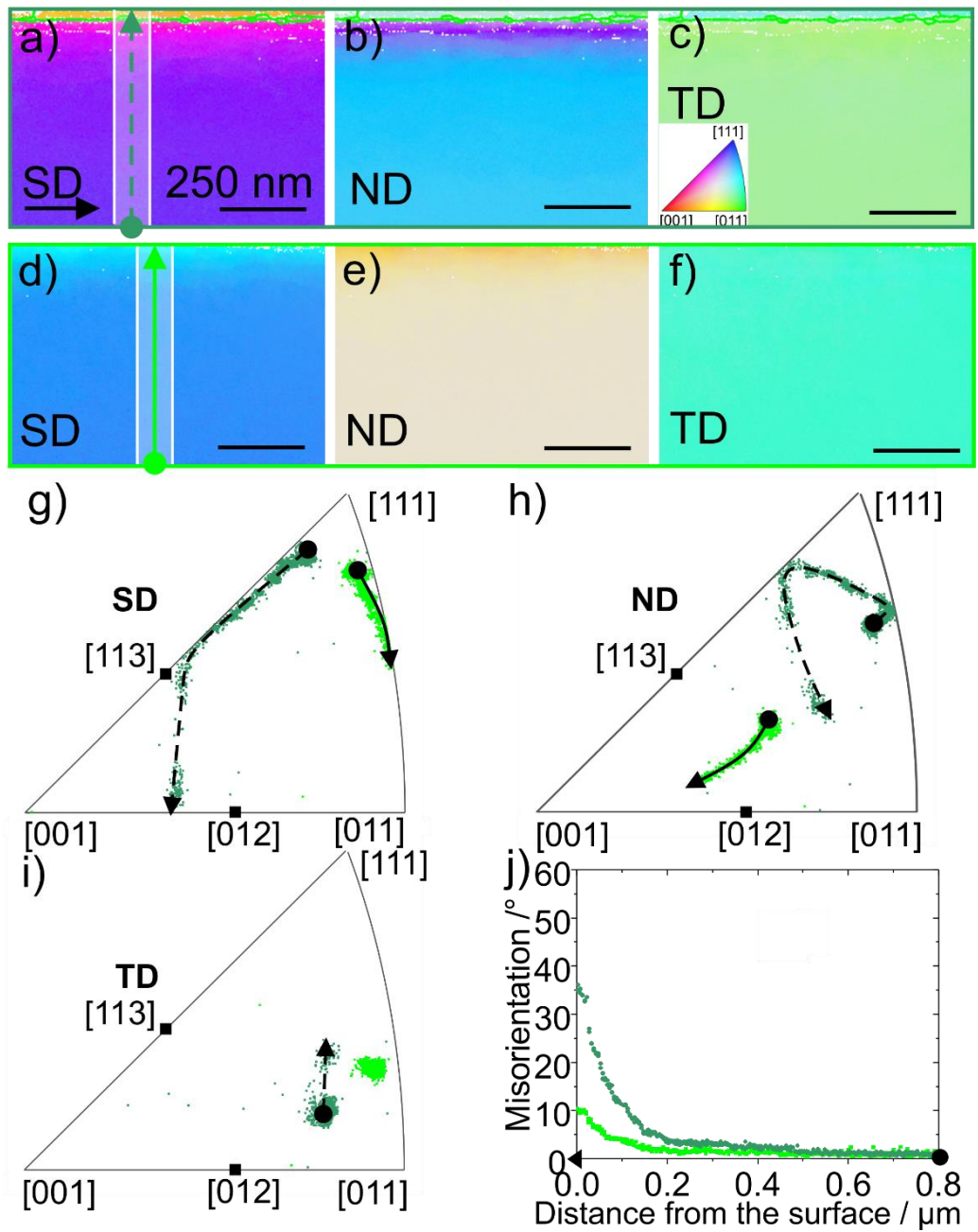

**Figure 4. HRTEM image of the line-like characteristics presented in Figure 2b.** In the areas marked by white squares, FFT analyses were carried out. In the FFT, the twins are indicated by dashed red lines and the matrix by dashed blue lines. The sliding direction was from left to right.

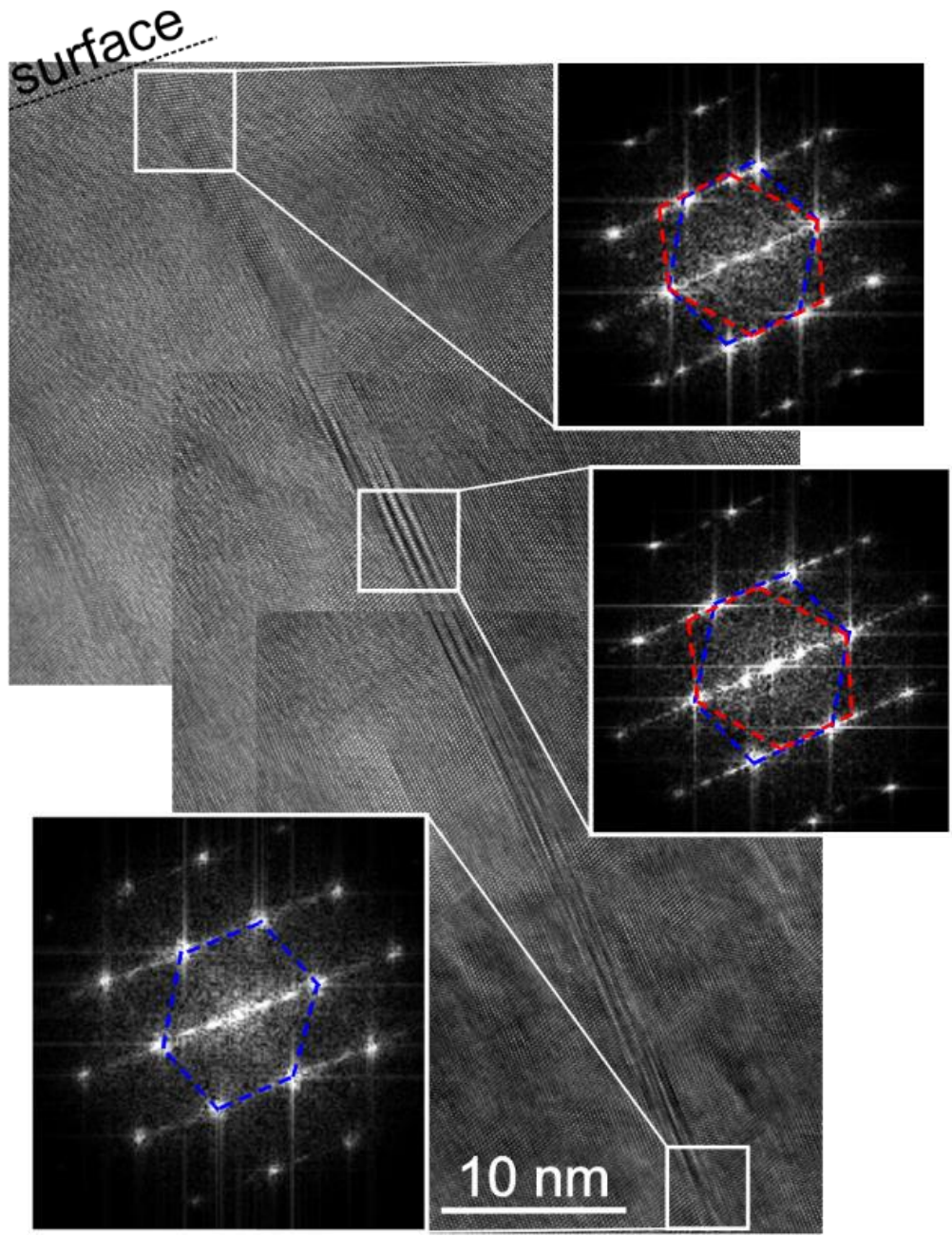

**Figure 5. Dislocation trace line and lattice rotation.** a) to c) TKD measurement of the SiC/2N/N$_2$ experiment (STEM in Figure 2c) and in d) to f) the SiC/5N/air experiment (STEM in Figure 2f) in SD, ND and TD, respectively. The colour-coding is given in the inset in c). Small-angle grain boundaries (3-15°) are highlighted in green, high-angle grain boundaries (>15°) in red and twins (Σ3) in blue. The SD is from left to right. g) misorientation profile along the arrows in a) and d). The distance 0 µm marks the surface.

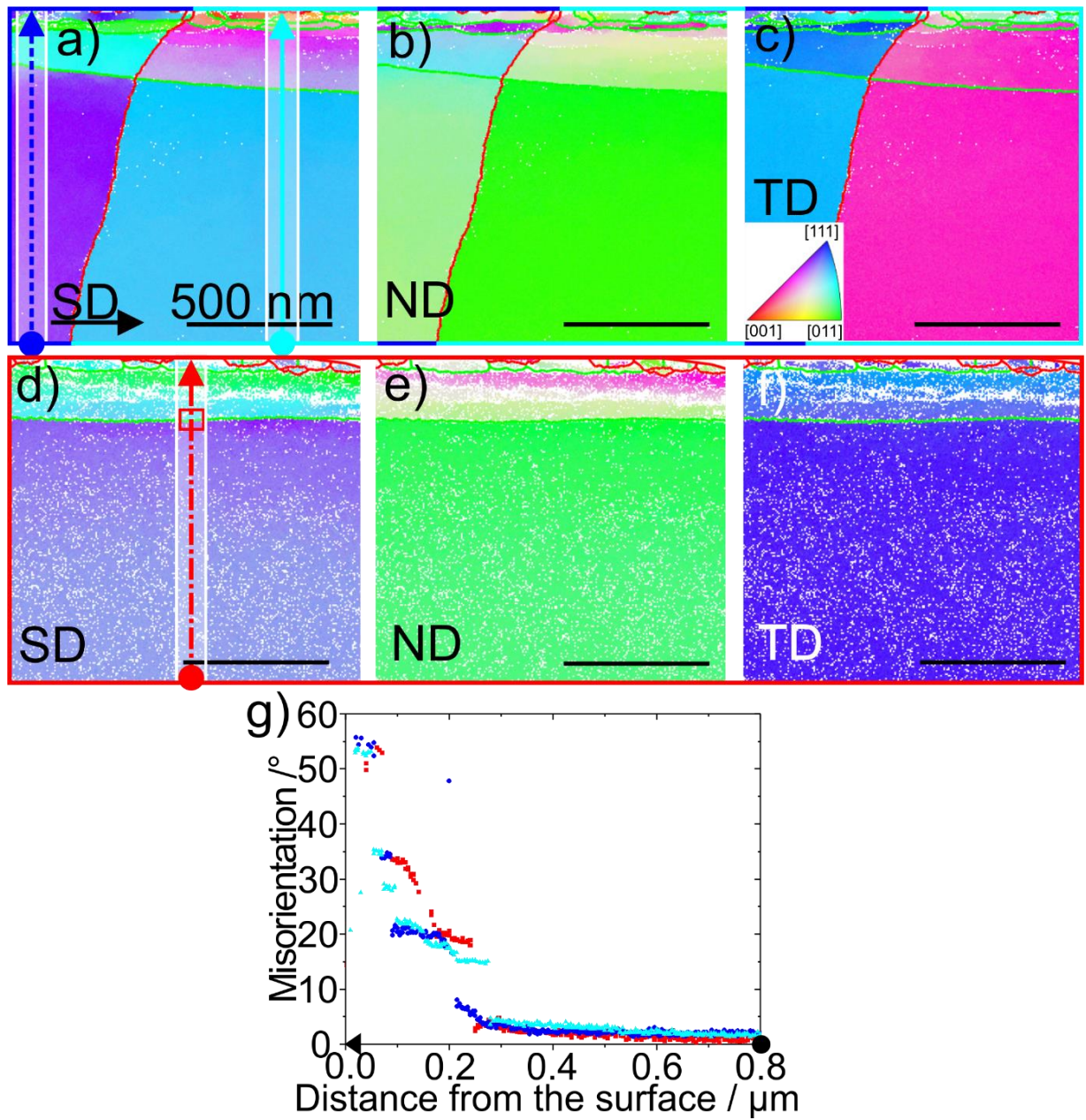

**Figure 6. Types of twin appearances.** a) to c) TKD measurement of the experiment for sapphire/2N/N$_2$ (STEM in Figure 2e) and d) to f) Si$_3$N$_4$/2N/air (STEM in Figure 2d). The measurements are colour-coded in SD, ND and TD. The colour-coding is given in the inset in c). Small-angle grain boundaries (3-15°) are highlighted in green, high-angle grain boundaries (>15°) in red and twins (Σ3) in blue. The sliding direction is from left to right.

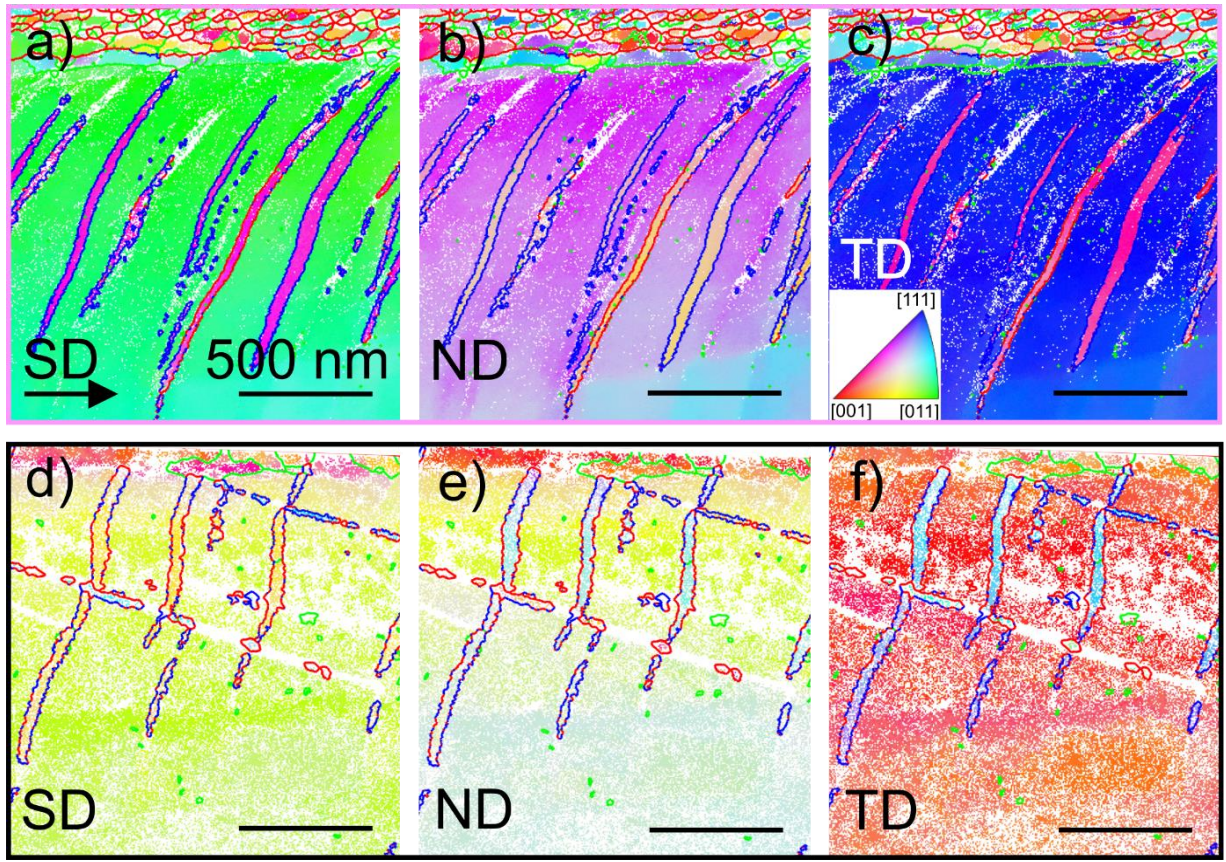

**Supplementary information**

**Figure S1. Images of the wear tracks and counter bodies after single trace experiments with a velocity of 0.5 mm/s on CoCrFeMnNi.** The first column shows the wear tracks, and the second one shows the spheres. The following systems are shown: a)SiC/2N/$N_2$, b)SiC/2N/air, 50%RH, c) $Si_3N_4$/2N/air, 50 %RH, d) sapphire/2N/$N_2$, e) sapphire/2N/air, 50%RH [7]and f) SiC/5N/$N_2$. The scale bar for all the images is given in a.

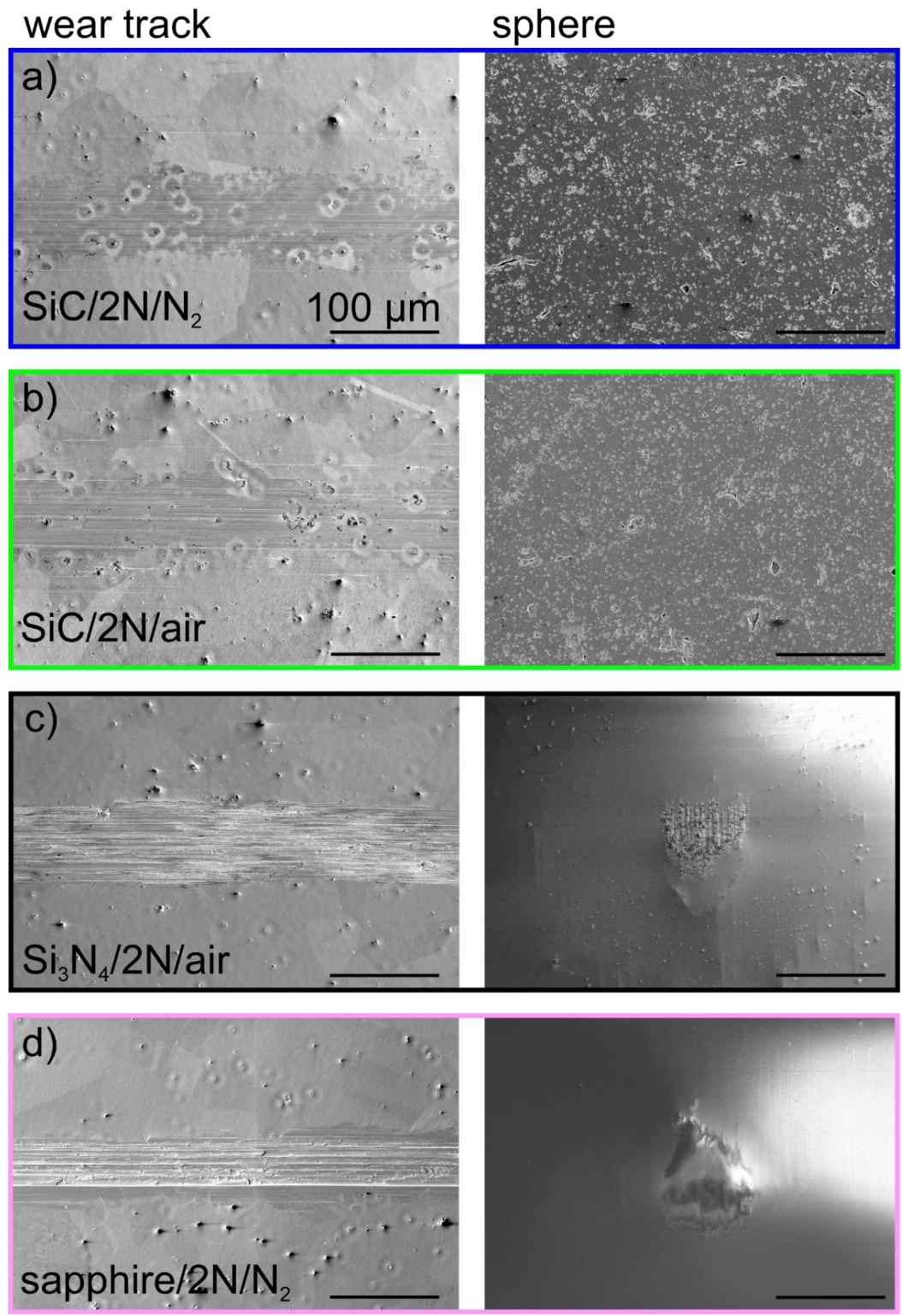

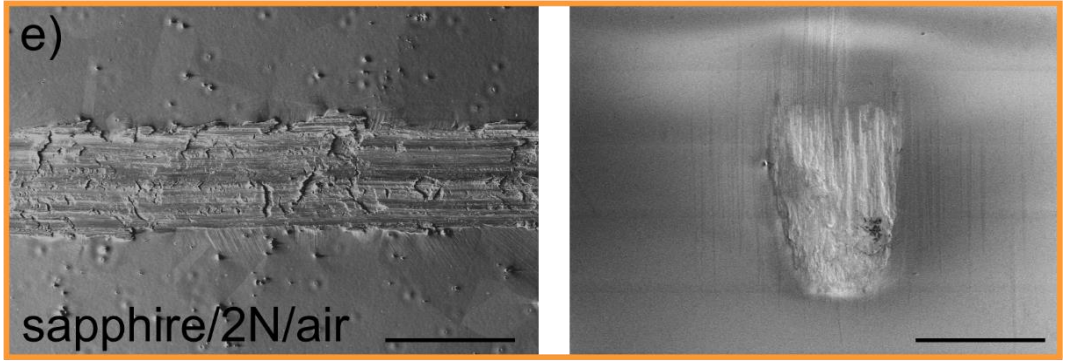

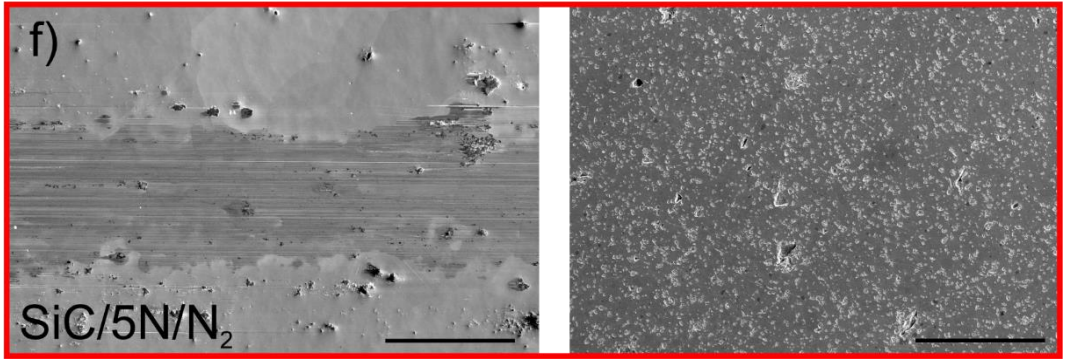

**Figure S2.** a) STEM image of the microstructure for the SiC/5N/air experiment within another grain compared to Figure 2f, b) to d) TKD measurement colour-coded in SD, ND and TD, respectively. The colour-coding is given in the inset in d). Small-angle grain boundaries (3-15°) are highlighted in green, high-angle grain boundaries (>15°) in red and twins (Σ3) in blue. The sliding direction is from left to right.

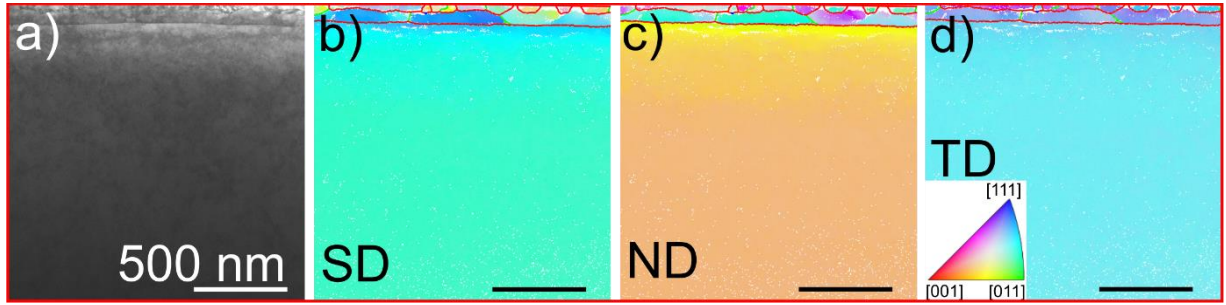

**Figure S3.** The orientation data in in the white boxes in Figure 5a+d were plotted in a) to c) in inverse pole figures (IPFs) in SD, ND and TD, respectively. The arrows within the IPFs point towards the surface. The square in a) is at the position of the DTL and also marked in Figure 5d.

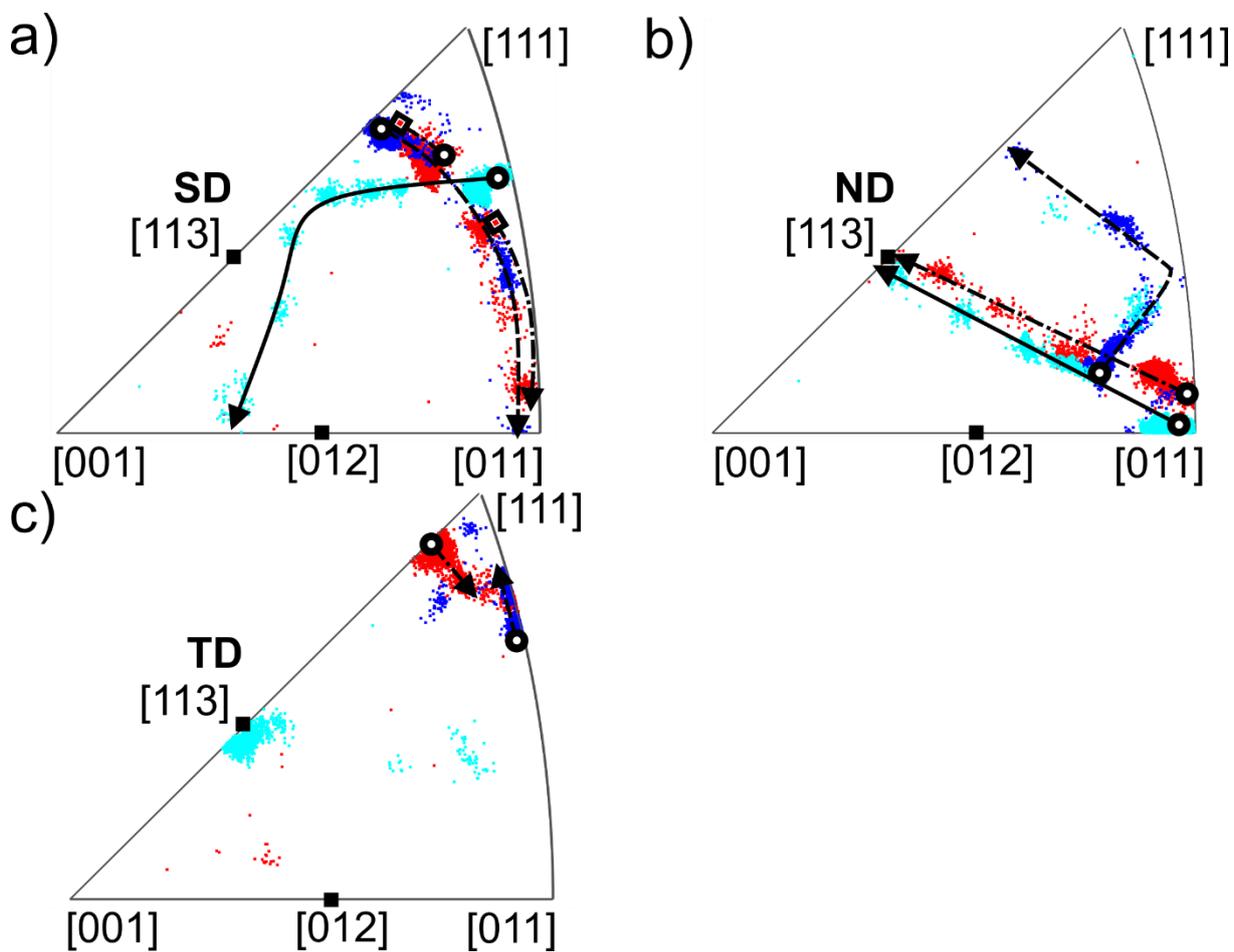

**Figure S4.** a) STEM image of the microstructure for the Si$_3$N$_4$/2N/air experiment within another grain compared to Figure 2d, b) to d) TKD measurement colour-coded in SD, ND and TD, respectively. The colour-coding is given in the inset in d). Small-angle grain boundaries (3-15°) are highlighted in green, high-angle grain boundaries (>15°) in red and twins (Σ3) in blue. The sliding direction is from left to right.

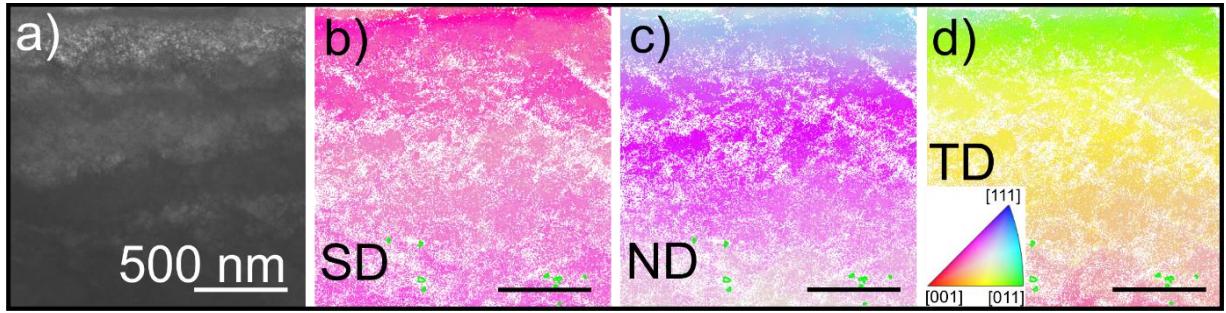

**Figure S5** Cross-sectional view of a molecular dynamics simulation with $Cu_{95}Ni_5$ after a) 0.0 ps, b) 779.2 ps, c) 781.6 ps, d) 797.0 ps, e) 797.6 ps and f) 804.2ps. a) shows the initial microstructure. Stacking faults (marked by black arrows) growing from the surface are observed in b) and c). d) to f) shows the twin thickening (marked by black arrows). The system from Ref. [47] was analysed at high time resolution using Ovito [50]. Fcc atoms are coloured in green, hcp in red and not assignable atoms in white. Black lines mark Shockley partial dislocations.

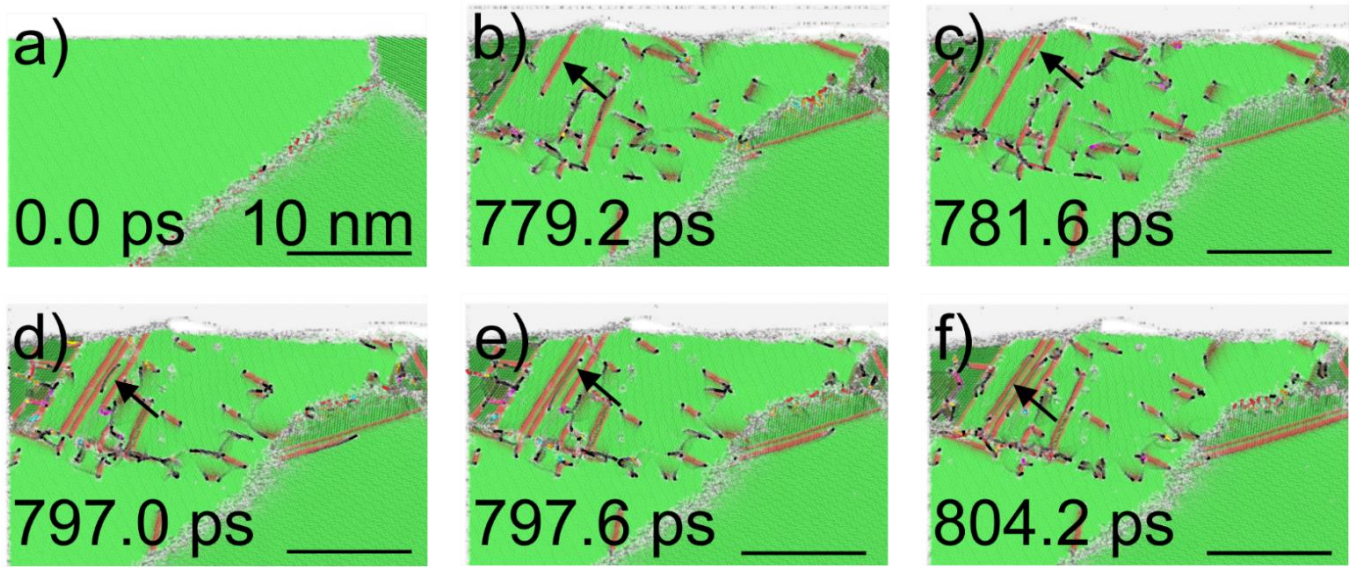